\pdfoutput=1
\documentclass[%
 reprint,
nofootinbib,
 amsmath,amssymb,
 aps,
]{revtex4-2}

\usepackage{graphicx}
\usepackage{dcolumn}
\usepackage{bm}

\usepackage{amsmath,amssymb,amsbsy,amstext,amsthm,simplewick,amsfonts,graphicx}
\usepackage{mathrsfs}
\usepackage{graphicx}
\usepackage{wrapfig}
\usepackage{upgreek}
\usepackage{bm} 
\usepackage{booktabs}
\usepackage{framed}
\usepackage{bbm}
\usepackage{textcomp}
\usepackage{adjustbox}
\usepackage{makecell}
\usepackage{tcolorbox}
\usepackage{empheq}
\usepackage[normalem]{ulem}
\usepackage{enumitem}
\usepackage{braket} 
\usepackage{array}
\usepackage{dsfont}
\usepackage{physics}
\usepackage{ulem}
\usepackage{xcolor}
\usepackage{tocloft}
\usepackage{tikz}
\usepackage{xcolor}
\usetikzlibrary{patterns}

\usepackage{hyperref}
\hypersetup{colorlinks=true,
	breaklinks=true,
	pdfstartview=Fit,
	linkcolor=blue2,
	citecolor=red2,
	urlcolor=blue
}

\definecolor{red2}{RGB}{214, 39, 40}
\definecolor{green2}{RGB}{0,170,0}
\definecolor{blue2}{RGB}{0,100,200}
\definecolor{magenta2}{RGB}{191,64,191}
\definecolor{purple2}{RGB}{112,48,160}
\definecolor{orange2}{RGB}{255,192,0}
\definecolor{green3}{RGB}{44,160,44}

\begin{document}

\title{\textit{Natura Non Facit Saltum}: \\An Analytical Model of Smooth Slow-Roll to Ultra-Slow-Roll Transition }
\author{Diego Cruces$^{1}$}
\email{dcruces@itp.ac.cn}
\author{Minxi He$^{2}$}
\email{heminxi@ibs.re.kr}
\author{Shi Pi$^{1,3}$}
\email{shi.pi@itp.ac.cn}
\author{Jianing Wang$^{3}$}
\email{jianing.wang@ipmu.jp}
\author{Masahide Yamaguchi$^{4,5,6}$}
\email{gucci@ibs.re.kr}
\author{Yuhang Zhu$^{4}$}
\email{yhzhu@ibs.re.kr}

\affiliation{
$^1$ Institute of Theoretical Physics, Chinese Academy of Sciences, Beijing 100190, China\\
$^2$ Particle Theory and Cosmology Group, Center for Theoretical Physics of the Universe, Institute for Basic Science (IBS), Daejeon, 34126, Korea\\
$^3$ Kavli Institute for the Physics and Mathematics of the Universe (WPI), UTIAS, The University of Tokyo, Kashiwa, Chiba 277-8583, Japan\\
$^4$ Cosmology, Gravity, and Astroparticle Physics Group, Center for Theoretical Physics of the Universe, Institute for Basic Science (IBS), Daejeon, 34126, Korea\\
$^5$ Department of Physics, Institute of Science Tokyo, 2-12-1 Ookayama, Meguro-ku, Tokyo 152-8551, Japan\\
$^6$ Department of Physics and IPAP, Yonsei University, 50 Yonsei-ro, Seodaemun-gu, Seoul 03722, Korea
}

\date{\today}

\begin{abstract}
In this \textit{letter}, we propose a single-field inflation model that realizes a slow-roll-to-ultra-slow-roll 
transition while keeping the second slow-roll parameter smoothly varying throughout.
The model is built through a minimal modification by introducing a simple time dependence in the effective mass term of the Mukhanov-Sasaki equation.  We obtain fully analytical solutions for both the background evolution and the curvature perturbations, which makes the parameter dependence of the curvature power spectrum easy to track. To the best of our knowledge, this is the first analytical model that describes a \textit{smooth} transition of this kind. We also compare its signatures with those of the corresponding sharp-transition counterpart.

\end{abstract}

\maketitle

\noindent \textbf{\emph{Introduction}---} One of the most profound ideas in modern cosmology is that the structure of the Universe we observe today can be traced back to quantum fluctuations in the very early period, most commonly described within the inflationary paradigm \cite{Starobinsky:1980te,Sato:1980yn,Guth:1980zm}. 
The anisotropies of the cosmic microwave background (CMB) and the formation of large-scale structure 
are among its best known imprints. On much smaller scales, primordial fluctuations may also give rise to primordial black holes (PBHs) \cite{Zeldovich:1967lct,Hawking:1971ei,Carr:1974nx,Carr:1975qj} as well as induced gravitational waves (IGWs) \cite{Tomita:1967wkp,Matarrese:1992rp,Matarrese:1993zf,Matarrese:1997ay,Noh:2004bc,Carbone:2004iv,Nakamura:2004rm,Ananda:2006af,Osano:2006ew,Baumann:2007zm,Saito:2008jc}, 
both of which have important implications for dark matter, super-massive black holes, GW events observed by LIGO--Virgo--KAGRA, stochastic GW background, energetic cosmic rays, and Hawking radiation \cite{Barrau:2001ev,Bean:2002kx,Duechting:2004dk,Kearsley:2006yp,Carr:2009jm,Kawasaki:2012kn,Sasaki:2016jop,Bird:2016dcv,Ali-Haimoud:2017rtz,Espinosa:2018euj,Bartolo:2018rku,Carr:2018rid,Caprini:2018mtu,Carr:2020gox,sasaki2025unveilingprimordialblackhole,petrov2026boundsgravitationalwaveproduction,Das:2021wei,He:2022wwy,He:2024wvt}. 
A significant amount of PBHs and IGWs can be produced if large curvature perturbations are abundantly generated during inflation \cite{Di:2017ndc,Cai:2018dig,Bartolo:2018evs,Unal:2018yaa,Ragavendra:2020sop,Pi:2020otn,Domenech:2020kqm,Wang:2021kbh,Pi:2021dft,Barker:2024mpz,Escriv__2024,Figueroa:2020jkf,Bhaumik:2020dor,Biagetti:2021eep,Kitajima:2021fpq,Geller:2022nkr,Cai:2022erk,Choudhury:2023hfm,Chen:2022dah,Firouzjahi:2023lzg,Frosina:2023nxu,Bhattacharya:2023ysp,Tada:2023rgp,LISACosmologyWorkingGroup:2023njw,Qin:2023lgo,Pi:2024ert,Domenech:2024rks,Croney:2024xzw,Firouzjahi:2025ihn,Sasaki:2025zao}, which can be realized when inflation has experienced an intermediate ultra-slow-roll (USR) phase~\cite{Tsamis:2003px,Kinney:2005vj,Martin:2012pe,Motohashi:2014ppa,Germani:2017bcs,Ballesteros:2017fsr,Liu:2020oqe,Cheng:2021lif,Figueroa:2021zah,Cheng:2023ikq,Hertzberg:2017dkh,Pattison:2018bct,Ballesteros:2020sre,Vennin:2020kng,Pattison:2021oen,Karam:2022nym,Pi:2022ysn,Firouzjahi:2023bkt,Kawaguchi:2024rsv,Fujita:2025imc,Franciolini:2023agm,Choudhury:2023hvf,Mishra:2023lhe,Ballesteros:2024zdp,Artigas:2024ajh,Inui:2024sce,Escriva:2025ftp}. 

To obtain reliable predictions from USR inflation without introducing spurious artifacts, the transition between slow-roll (SR) and USR phases must be handled carefully.
In most existing studies, however, 
the second SR parameter $\epsilon_2$ jumps discontinuously at the SR-USR transition \cite{Yi:2017mxs,Morse:2018kda,Carrilho:2019oqg,Motohashi:2019rhu,Choudhury:2023jlt,Liu:2020oqe,Motohashi:2023syh}. Such a discontinuity is physically unnatural and enters directly into the equation of motion for the curvature perturbation. 
For instance, a discontinuous transition leads to instantaneous mode mixing in the curvature perturbations, as the Mukhanov-Sasaki (MS) equation is solved separately on the two sides of the transition. The resulting effect is therefore an artifact of the abrupt transition, rather than a genuine consequence of the underlying physical dynamics.

In contrast, the transition can proceed smoothly, as in models with an inflection-point potential \cite{Germani:2017bcs,Bhaumik:2019tvl,Tomberg:2025fku} or multi-field inflation scenarios \cite{Geller:2022nkr}. A smooth transition avoids spurious effects, retaining only the physical features. 
The evolution of both background and perturbations in smooth-transition models has been studied numerically~\cite{Germani:2017bcs,Motohashi:2017kbs,Kannike:2017bxn,Ballesteros:2017fsr,Bhatt:2022mmn,Caravano:2024moy,Caravano:2025diq,Allegrini:2025jha}. These studies show that a smooth transition can leave distinct imprints on the curvature power spectrum~\cite{Byrnes:2018txb,Inomata:2018cht,Passaglia:2018ixg}. 
In particular, the amplitude and tilt of the spectrum are determined by the time evolution of the second slow-roll parameter, the inflaton velocity, and the duration of the USR phase \cite{Tasinato:2020vdk,Ezquiaga:2018gbw,Franciolini:2022pav}. Further developments along this direction include numerical computation of bispectrum \cite{Adshead:2013zfa,Ragavendra:2023ret,Ragavendra:2020sop}, stochastic inflation approach~\cite{Cruces:2018cvq,Pattison:2019hef,Pattison:2021oen,Sharma:2024fbr}, and multi-field embeddings~\cite{Geller:2022nkr,Su:2025mam}.

Despite extensive works on USR inflation, an analytically solvable USR model with a smooth SR-USR transition is still missing. Previous studies address at most one of them: either analytically tractable \cite{Carrilho:2019oqg,Pi:2022zxs,Karam:2022nym,Tada:2023rgp,Firouzjahi:2023bkt,Motohashi:2023syh,Maity:2023qzw,Kawaguchi:2024rsv,Fujita:2025imc} or smoothly evolving \cite{Germani:2017bcs,Ballesteros:2017fsr,Di:2017ndc,Ezquiaga:2018gbw,Passaglia:2018ixg,Cai:2018dkf,Byrnes:2018txb,Dalianis:2018frf,Dalianis:2018frf,Bhaumik:2019tvl,Ragavendra:2020sop,Liu:2020oqe,Ballesteros:2020qam,Iacconi:2021ltm,Cheng:2021lif,Figueroa:2021zah,Wang:2021kbh,Figueroa:2021zah,Iacconi:2021ltm,Bhatt:2022mmn,Cole:2022xqc,Ragavendra:2023ret,Jackson:2023obv,Cheng:2023ikq,Davies:2023hhn,Barker:2024mpz,Sharma:2024fbr,Jackson:2024aoo,Caravano:2024moy,Caravano:2025diq,Briaud:2025hra}. 
The aim of this paper is to fill this gap by constructing a model that satisfies both requirements. Specifically, we consider the inflationary scenario with a three-stage evolution: an SR phase (SR1) followed by a USR phase, which finally ends with another SR phase (SR2). \textit{Nature does not make jumps}, so we require the transitions among these phases to be \textit{smooth}. 
Simultaneously, we aim for an \textit{analytically} solvable model, in the sense that both the slow-roll parameters and the primordial power spectrum can be expressed in terms of elementary or special mathematical functions. Such a setup will not only simplify practical applications but also make the underlying physics more transparent.

\vspace{0.5cm}
\noindent \textbf{\emph{The Model}---} Here, we specify the criteria of our construction to make the model both \textit{analytic} and \textit{smooth}: 
\begin{enumerate}[label=(\roman*), ref=(\roman*)]
	\item The second slow-roll parameter $\epsilon_2$
 evolves as a continuously differentiable function ($\epsilon_2 \in C^1$), from the first SR stage compatible with CMB constraints (i.e. with $\epsilon_2\simeq0$) into a USR regime.\label{it:condition2}
	\item $\epsilon_2$ exits the USR regime with a bounded and continuous first derivative and reaches an attractor solution, thereby allowing inflation to end.\label{it:condition3}
	\item Both slow-roll parameters and the power spectrum of curvature perturbation admit analytical expressions.  \label{it:condition4}
\end{enumerate}
	Throughout the three stages, the MS variable $v_k$ always satisfies the following equation \cite{Mukhanov:1985rz,Sasaki:1986hm}
\begin{equation}
		v_k^{\prime \prime}+\left(k^2-\frac{\nu^2-1/4}{\tau^2}\right) v_k=0\,, \label{eq: MS_equation}
\end{equation}
where the effective index $\nu$ is defined through $\nu^2\equiv {1}/{4}+({z''}/{z})\,\tau^2$, and $z\equiv (a {\phi'})/{\mathcal{H}}$ is determined by the evolution of the inflaton background $\phi$. All the primes represent derivatives with respect to the conformal time $\tau$. $\mathcal{H}\equiv a'/a$ is the conformal Hubble parameter. In general, $\nu$ can be an arbitrary time-dependent function, depending on the inflaton dynamics. 

Assuming the first slow-roll parameter $\epsilon_1\equiv 1-\mathcal{H}'/\mathcal{H}^2$ is always negligible before the end of inflation (will be verified later), $\nu$ can be expressed as \cite{Stewart:1993bc,Chen:2006nt,Kristiano_2025}
\begin{equation}
		\nu^2 = \frac{9}{4}+\frac{3}{2}\epsilon_2+\frac{1}{4}\epsilon_2^2+\frac{1}{2}\frac{\epsilon_2'}{\mathcal{H}}\equiv\frac{9}{4}+\mathcal{F}(\tau)\,,
		\label{nu_epsilon_2}
\end{equation}
where the second slow-roll parameter $\epsilon_2 \equiv \epsilon_1'/(\mathcal{H}\epsilon_1)$ and 
$\mathcal{F}(\tau)$ encodes the time dependence. 
$\epsilon_2$ characterizes the rate of change of 
$ \epsilon_1 $, which is closely related to the amplitude of the curvature perturbation power spectrum. 
During SR, $|\epsilon_2| \ll 1$, 
$ \epsilon_1 $ is almost constant and therefore 
the curvature perturbation stope evolving on superhorizon scales. 
However, $\epsilon_2$ becomes negatively large during the USR phase, causing a rapid decay of $ \epsilon_1 $ 
and consequently the growth of the curvature perturbation. 
Consistency with CMB constraints requires an SR phase at CMB scale, during which $\epsilon_2$ remains small and $\nu \simeq 3/2$. 

\begin{figure}[t]
\centering
\includegraphics[width=0.48\textwidth]{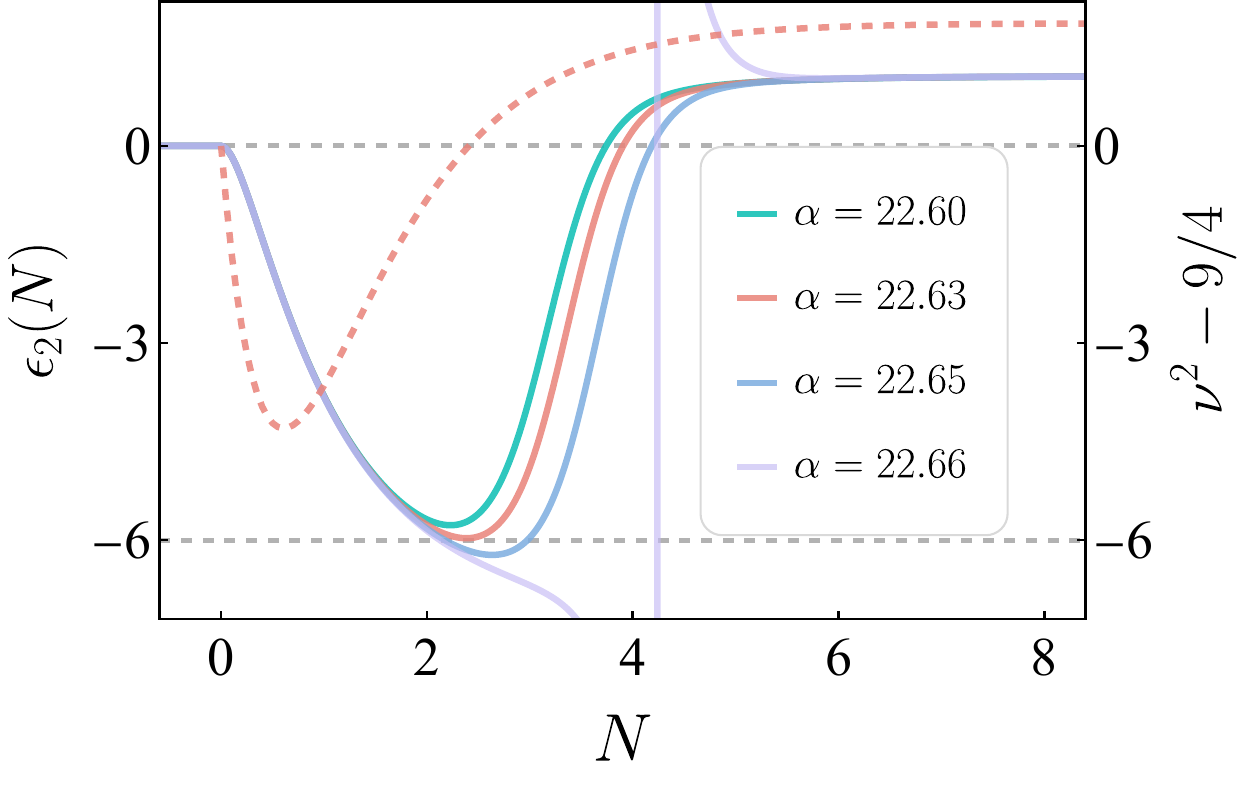}
\caption{The evolution of $\epsilon_2(N)$ with different choices of the parameter $\alpha$, while we fix the combination $\left(\mu^2-9/4\right)/q^2=0.09$. $\nu^2(N)-9/4$ with $\alpha=22.63$ and $\mu=2.0294$ is shown as the \textcolor[RGB]{236,149,140}{red dashed} line.}
\label{fig:diffBs}
\end{figure}

To achieve our goal \ref{it:condition4}, Eq.~\eqref{eq: MS_equation} must admit an analytical solution, 
while the index $\nu$ should be simple enough.
The simplest choice of a constant $\nu^2>0$, as is commonly adopted in the literature \cite{Motohashi:2014ppa,Anguelova:2017djf,Motohashi:2017aob,Odintsov:2017qpp,Pattison:2019hef,Guerrero:2020lng,Ahmadi:2023qcw,Inui:2024sce,Ballesteros:2024zdp,Kristiano:2024vst,Briaud:2025hra,Fujita:2025imc,Motohashi:2025qgd}, 
can satisfy \ref{it:condition3} and \ref{it:condition2} simultaneously. 
This is because the solution for $\nu^2=\text{constant}>0$ in Eq.~\eqref{nu_epsilon_2} is given by hyperbolic functions that have two asymptotic values at $ \tau \to -\infty $ and $ \tau \to 0 $ with $\epsilon_2(\tau \to -\infty)\leq \epsilon_2(\tau \to 0)$. 
Given that $\epsilon_2^{\rm{USR}}<\epsilon_2^{\rm{SR}}$, it is impossible to match it with SR on both sides. 
If $\nu^2=\text{constant}<0$~\cite{Ballesteros:2024zdp}, the solution is given by tangent functions which are not desirable for our purpose. 

The next simplest yet sufficiently general option is a polynomial function of time, 
\begin{align}
	\mathcal{F}(\tau)=\left(\mu^2-\frac{9}{4}\right)-\alpha\left(\frac{\tau}{\tau_\star}\right)+q^2\left(\frac{\tau}{\tau_\star}\right)^2\,,\label{eq: functionF}
\end{align}
which is the model we consider in this \textit{letter}. Here we keep only the first two powers of $\tau$, since higher power terms such as $\propto \tau^3$ modify the behavior of Eq.~\eqref{eq: MS_equation} as they will dominate over $k^2$ at the far past, while negative powers of $\tau$ alter the super-horizon behaviors. 
There are three independent 
dimensionless
parameters in our model, $\mu$, $\alpha$, and $q$, all of which are assumed to be positive. 
Continuity of $\nu$ at the SR1-USR transition time $\tau_{\star}$ requires 
$ \mathcal{F}(\tau_\star)=0 $, 
which reduces the number of independent parameters by one via the following constraint on the parameters
\begin{equation}
   \mu^2=\frac{9}{4}+\alpha-q^2\,.
   \label{condition_continuity}
\end{equation}
As we will see, the two independent parameters control how inflation ends and the duration of the non-attractor regime with $\epsilon_2<-3$.

As mentioned above, throughout this paper we will 
neglect $\mathcal{O}(\epsilon_1)$ corrections (see App.~\ref{apdx:dS} for details), meaning the Hubble parameter during inflation will be a constant $H_0$. With this simplification, Eq.~\eqref{nu_epsilon_2} can be rewritten in terms of the $e$-fold number $N$ as 
\begin{equation}
	\frac{3}{2}\epsilon_2+\frac{1}{4}\epsilon_2^2+\frac{1}{2}\frac{\mathrm{d}\epsilon_2}{\mathrm{d}N}=\left(\mu^2-\frac{9}{4}\right)-\alpha\,e^{-N}+q^2e^{-2 N}\,,
	\label{evol_epsilon2_N}
\end{equation}
where we have used $a=-1/{H_0 \tau}\,$ and  $N=-\ln\left(\tau/\tau_\star\right)$.
The differential equation 
Eq.~\eqref{evol_epsilon2_N} can be solved analytically for $\epsilon_2$, and the result is presented in App.~\ref{apdx:coeffi}. 
So far, we have achieved the first half of \ref{it:condition4}. 

We plot 
$\epsilon_2$ in Fig.~\ref{fig:diffBs} for several representative parameter choices. 
Clearly, $\epsilon_2$ evolves smoothly from its SR1 value ($\epsilon_2^{\rm{SR1}}\simeq 0$) to its final SR2 value ($\epsilon_{2,f}\simeq -3+2\mu$), with an intermediate USR (or non-attractor) phase where $-3>\epsilon_2>-3-2\mu$,  satisfying in this way our conditions \ref{it:condition2} and \ref{it:condition3}. In Fig.~\ref{fig:diffBs} we can also see that if we choose parameters such that $\epsilon_2<-3-2\mu$ at some point in its evolution (\textcolor[RGB]{226,220,249}{purple line}), it becomes discontinuous and hence not desirable for our purposes. For reference, we also plot the corresponding function $\nu^2(N)$.
\begin{figure}[t]
\centering
\hspace{-0.5cm}
\includegraphics[width=0.45\textwidth]
{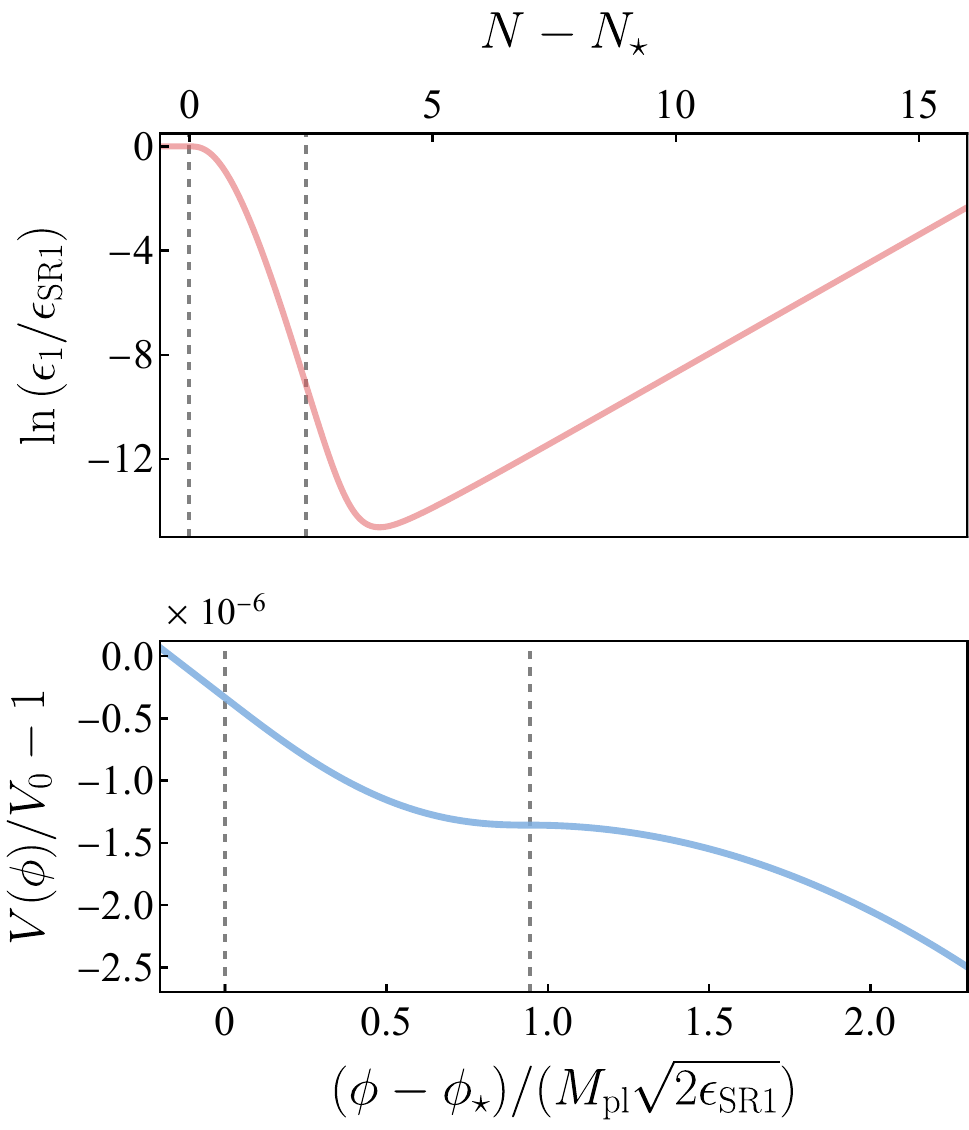}
\caption{\textit{Top}: The evolution of the leftmost slow-roll parameter $\epsilon_1(N)$. $\textit{Bottom}$: The inflaton potential $V(\phi)$ of our model. The first dashed line (the origin of the horizontal axis) marks the time $\tau_\star$ and the other one indicates when $\epsilon_2$ reaches value $-6$. By setting $\epsilon_2=-6$ and $\mathrm{d}\epsilon_2 /\mathrm{d} N=0 $ in Eq.~\eqref{evol_epsilon2_N}, the $e$-fold number corresponding to $\epsilon_2=-6$ is given as $N=-\ln[\alpha/(\alpha-\mu^2+9/4)-1]$. The parameters used in these plots are $\alpha=22.63$ and $\mu=2.0294$ and $\epsilon_{\text{SR1}}=10^{-6}$.  }
\label{fig:Vphi}
\end{figure}

\vspace{0.5cm}
\noindent \textbf{\emph{Power Spectrum and Features}---} To fully achieve \ref{it:condition4}, we need to solve the mode function $ v_k(\tau) $ and function $ z(\tau) $ analytically to obtain the curvature perturbation $ \mathcal{R}_k=v_k/z $.  

By inserting Eq.~\eqref{eq: functionF} into Eq.~\eqref{eq: MS_equation}, $ v_k(\tau) $ admits a general solution in terms of the Whittaker-$M$ and $W$ functions: 
\begin{equation}
v_k(\tau)=\mathcal{C}_1\cdot W_{i\widetilde{\alpha},\mu}\left( 2 i k_{\text{eff}} \tau\right)+\mathcal{C}_2\cdot M_{i\widetilde{\alpha},\mu}\left( 2 i k_{\text{eff}} \tau\right)\,,\label{eq:solution_of_v}
\end{equation}
where we introduced $k_\star\equiv -1/\tau_\star$, 
$k^2_{\text{eff}}\equiv k^2-(q k_\star)^2$, and $\widetilde{\alpha}\equiv \alpha k_\star/2k_{\text{eff}}$. 
The coefficients $\mathcal{C}_{1,2}$ will be determined later by appropriate boundary conditions.

According to Eqs.~\eqref{nu_epsilon_2} and \eqref{eq: functionF}, function $ z(\tau) $ satisfies 
\begin{equation}
\frac{1}{4}+\frac{z''}{z}\tau^2=\mu^2-\alpha\left(\frac{\tau}{\tau_\star}\right)+q^2\left(\frac{\tau}{\tau_\star}\right)^2\,. \label{eq:zppz}
\end{equation}
This equation can also be solved 
analytically 
and expressed by the same types of special functions as 
\begin{equation}
z(\tau)=\mathcal{B}_1\cdot W_{\alpha/2q,\mu}\left( 2 q {\tau}/{\tau_\star}\right)+ \mathcal{B}_2\cdot M_{\alpha/2q,\mu}\left( 2 q {\tau}/{\tau_\star}\right)\,. \label{eq:solution_of_z}
\end{equation}
Given the relation $z(\tau)\approx- \sqrt{2 \epsilon_1(\tau)}M_{\rm{pl}}/(H_0 \tau)$, the coefficients $\mathcal{B}_{1,2}$ can be determined by boundary conditions $\epsilon_1|_{\tau_\star}=\epsilon_{\mathrm{SR}1}$, $\epsilon_1'|_{\tau_\star}=0$, which are imposed by smooth connection between SR1 and USR. 
Here $\epsilon_{\mathrm{SR}1}$ denotes the first slow-roll parameter during SR1 that is related to the amplitude of CMB-measured power spectrum of curvature perturbation via $\mathcal{P}^{\text{SR1}}_{\mathcal{R}}=H^2/(8\pi^2\epsilon_{\mathrm{SR1}}M_{\text{pl}}^2)$. 
The explicit expressions for the coefficients are collected in App.~\ref{apdx:coeffi}. 
The analytical result of $\epsilon_1(N)$ follows right after fixing the exact form of $z(\tau)$ in Eq.~\eqref{eq:solution_of_z}. The corresponding potential can be reconstructed using $V=M_{\mathrm{Pl}}^2 H^2 \left(3-\epsilon_1\right)$, with $H=H_0 \exp\left(-\int \epsilon_1\mathrm{d}N\right)$. The results of $\epsilon_1$ and $V(\phi)$ are presented in Fig.~\ref{fig:Vphi}.

\begin{figure}
\centering
\includegraphics[width=0.45\textwidth]{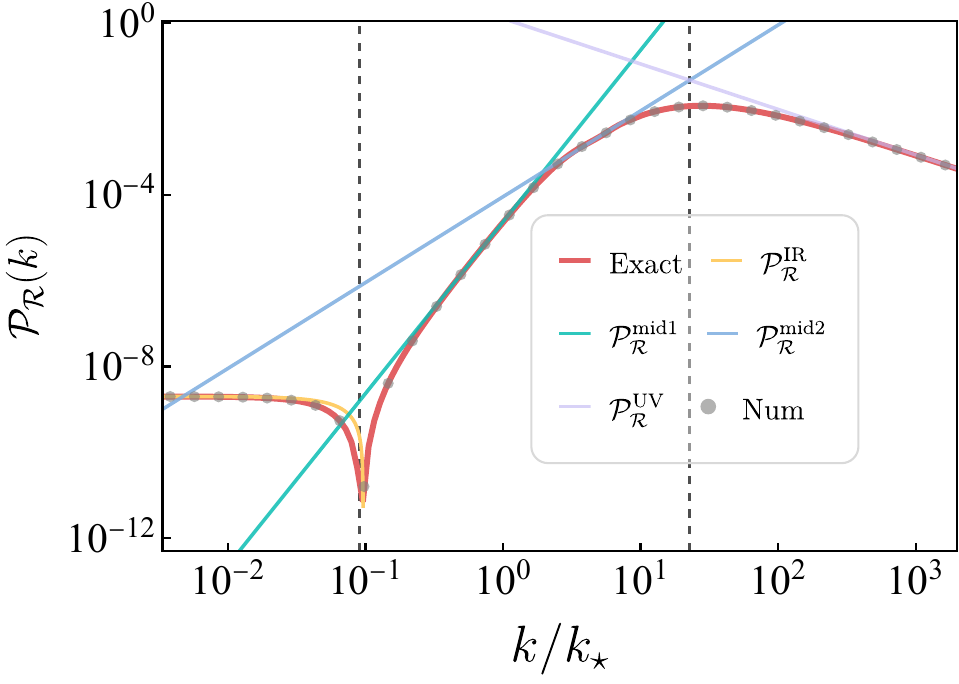}
\caption{Comparison of the analytically derived power spectrum (\textcolor[RGB]{227,97,100}{red line}) with the approximation formulae in different regions (\textcolor[RGB]{255,205,104}{yellow} for the IR region, \textcolor[RGB]{10,190,180}{green} and \textcolor[RGB]{125,174,224}{blue} for the growth region, \textcolor[RGB]{197,187,245}{purple} for the UV region), as well as with the numerical results (\textcolor[RGB]{178,179,178}{gray dots}). The dip and the peak location are shown by the black dashed lines. The parameters are set to $\tau_{\star}=-1$, $\alpha=22.63$ and $\mu=2.0294$.} 
\label{fig:NumVSAnaVsApp}
\end{figure}

Combining Eqs.~\eqref{eq:solution_of_v} and \eqref{eq:solution_of_z}, we arrive at the following analytical expression for the curvature perturbation, 
\begin{align}
    \mathcal{R}_k=\frac{\mathcal{C}_1 W_{i\widetilde{\alpha},\mu}\left( 2 i k_{\text{eff}} \tau\right)+\mathcal{C}_2 M_{i\widetilde{\alpha},\mu}\left( 2 i k_{\text{eff}} \tau\right)}{\mathcal{B}_1 W_{\alpha/2q,\mu}\left( 2 q \tau / \tau_\star\right)+ \mathcal{B}_2 M_{\alpha/2q,\mu}\left( 2q\tau/ \tau_\star\right)}\,. \label{eq: curvatureR}
\end{align}
During SR1, it should recover the 
Bunch-Davies (BD) solution
\begin{align}
    \mathcal{R}^{\text{BD}}_k =  \frac{H}{M_{\rm{pl}}\sqrt{\epsilon_{\rm{SR1}}}}\frac{i}{2\sqrt{k^3}}(1+ik\tau)e^{-ik\tau}\,,
\end{align}
so our boundary conditions are 
\begin{align}
\mathcal{R}_k\big|_{\tau_\star}=\mathcal{R}_k^{\rm{BD}}\big|_{\tau_\star}\,,\quad \mathcal{R}'_k\big|_{\tau_\star}=\mathcal{R}_{k}^{\rm{BD}\prime}\big|_{\tau_\star}\,,
\end{align}
which pin down the coefficients $\mathcal{C}_{1,2}$.
The detailed expressions are again collected in App.~\ref{apdx:coeffi}.
In the limit $\tau \to 0^{-}$, the Whittaker-$M$ decays more rapidly than the Whittaker-$W$. As a result, when evaluating the power spectrum at the end of inflation  i.e. $\mathcal{P}_{\mathcal{R}}(k)\equiv (k^3/2\pi^2) |\mathcal{R}_k(0^-)|^2 $, it is sufficient to take Eq.~\eqref{eq: curvatureR} as
\begin{align}\label{eq:R_exact}
    \mathcal{R}_k(0^-)&=\lim_{\tau\to0^-}\frac{\mathcal{C}_1\cdot W_{i\widetilde{\alpha},\mu}\left( 2 i k_{\text{eff}} \tau\right)}{\mathcal{B}_1\cdot W_{\alpha/2q,\mu}\left( 2 q \tau / \tau_\star\right)}\,\nonumber\\
    &=\frac{\mathcal{C}_1}{\mathcal{B}_1} \frac{\Gamma\left(1/2+\mu-{\alpha}/{2 q }\right)}{\Gamma\left(1/2+\mu-{i \widetilde{\alpha}} \right)} \left( \frac{-i k_{\mathrm{eff}} }{q k_\star}\right)^{\frac12-\mu}\,.
\end{align} 
The analytical results are verified in Fig.~\ref{fig:NumVSAnaVsApp} by comparing them with the numerical results, demonstrating excellent agreement. 

These analytical expressions also allow the properties of the final power spectrum in different regions to be readily understood by taking the appropriate limits.
\begin{enumerate}
    \item[$\bullet$] \textit{Infrared (IR) region}. When $k\ll k_{\star}$, we perform a power series expansion around $k/k_\star\to 0$ as
    \begin{align} 
    \label{eq:PIR}
    \mathcal{P}_{\mathcal{R}}^{\mathrm{IR}}(k)
    =\frac{a_0+a_2 k^2}{2\pi^2 |\mathcal{B}_1|^2 }\,,
    \end{align}
    where $ a_0 $ and $ a_2 $ are specified in Appendix \ref{apdx:approx}, where one can check that they are of opposite sign. 
    Due to the sign difference, a dip appears in the power spectrum. Its location can be determined straightforwardly by solving for the zero of the above approximation as
    \begin{equation}\label{eq:Pkdip}
    k_{\mathrm{dip}}=\left(-\frac{a_0}{a_2}\right)^{1/2}.
    \end{equation}
    \item[$\bullet$] \textit{Growth region.} When $k\sim k_\star $, our model enters the region exhibiting the 
    $k^4$-enhancement \cite{Byrnes:2018txb,Ozsoy:2019lyy,Tasinato:2020vdk,Cole:2022xqc,Domenech:2023dxx,Jackson:2023obv,Artigas:2024ajh,Briaud:2025hra}. Using the dip location, the power spectrum around this region can be approximated as 
    \begin{align}\label{eq:Pmid1}
    \mathcal{P}_{\mathcal{R}}^{\text{mid1}}(k)=\mathcal{P}_{\mathcal{R}}^{\rm{SR1}}\,\left(\frac{k}{k_{\mathrm{dip}}}\right)^4\,.
    \end{align}
    However, around $k\sim q k_\star>k_{\star}$, we find that the infrared scaling law deviates from  $k^4$ and instead approaches to $k^2$. We approximate the power spectrum in this region as
    \begin{align}\label{eq:Pmid2}
    \mathcal{P}_{\mathcal{R}}^{\text{mid2}}(k )=\mathcal{P}_{\mathcal{R}}(q k_\star)\,\left(\frac{k}{q k_\star}\right)^2\,,
    \end{align}
    where the closed form for $\mathcal{P}_{\mathcal{R}}$ at $k=q k_\star$ can be found in Eq.~\eqref{eq:Psig}.
    \item[$\bullet$] \textit{Ultraviolet (UV) region}. For modes exiting the horizon at very late stages, in the limit $k\gg k_\star $, we have $|\mathcal{C}_1(k)|\sim 1/\sqrt{2k}$ and $k_{\rm{eff}}\sim k$. The dimensionless power spectrum is approximated by
    \begin{align}\label{eq:PUV}
    \mathcal{P}_{\mathcal{R}}^{\mathrm{UV}}(k)=\frac{k^2}{4\pi^2} \left( \frac{k}{q k_\star}\right)^{1-2\mu} \frac{\Gamma\left(1/2+\mu-{\alpha}/{2 q}\right)^2}{|\mathcal{B}_1|^2\,\Gamma\left(1/2+\mu\right)^2}\,.
    \end{align}
    Therefore, the UV scaling behaves as $k^{3-2\mu}$, and it becomes scale-invariant when $\mu=3/2$, as expected.
\end{enumerate}
Making use of these formulae, we can estimate the peak location by solving for the intersection point of Eq.~\eqref{eq:Pmid2} and Eq.~\eqref{eq:PUV}, which gives 
\begin{align}\label{eq:Pkpeak}
    k_{\rm{peak}}=\left[\frac{(qk_\star)^{2\mu+1}}{4\pi^2\mathcal{P}_{\mathcal{R}}(qk_\star)}  \frac{\Gamma\left(\frac12+\mu-{\alpha}/{2 q}\right)^2}{|\mathcal{B}_1|^2\,\Gamma\left(\frac12+\mu\right)^2}\right]^{\frac{1}{2\mu-1}}.
\end{align}
In Fig.~\ref{fig:NumVSAnaVsApp}, we compare the approximation formulae in different regions with the exact result from Eq.~\eqref{eq:R_exact}. The estimated locations of the dip and the peak, indicated by the black dashed lines, are obtained from Eq.~\eqref{eq:Pkdip} and Eq.~\eqref{eq:Pkpeak}, respectively. They are all in excellent agreement with the exact curve.

\begin{figure}[t]
\centering
\hspace{-1cm}
\includegraphics[width=0.45\textwidth]{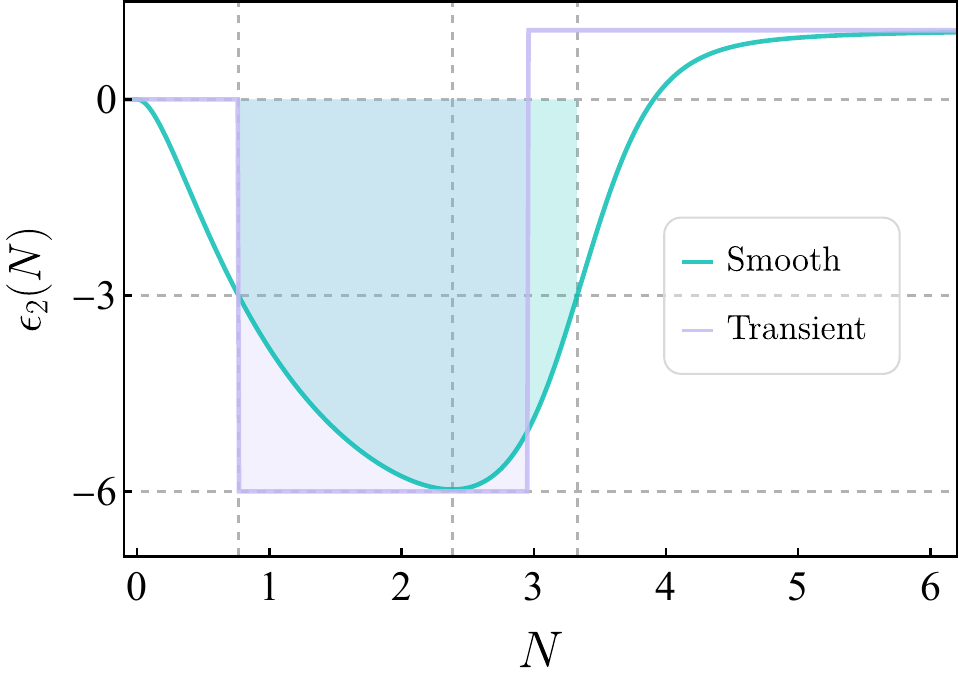}
\caption{The comparison of $\epsilon_2(N)$ between our smooth model (\textcolor[RGB]{10,190,180}{green line}) and the transient model (\textcolor[RGB]{197,187,245}{purple line}). For the  smooth one we take $\alpha=22.63$ and $\mu=2.0294$, while the transient one is specified by Eq.~\eqref{eq:sharpmodel}. The shaded regions correspond to the areas where the models enter the non-attractor phase.}
\label{fig:epsilon2_smoothVSsharp}
\end{figure}

\vspace{0.5cm}
\noindent\textbf{\emph{Comparison with Transient Model}---}
\label{s:4}
A natural next step is to examine how it contrasts with models characterized by sharp transitions.
In this section, we compare our model with its sharp counterpart which we refer to as the transient model. In this model, $\epsilon_2$ is defined by the following piecewise function,
\begin{align}\label{eq:sharpmodel}
    \epsilon_{2}(N)=
    \begin{cases}
        ~~~0\qquad &N<N_1\,,\\
    ~-6\qquad &N_1<N<N_1+N_{\text{eff}}\,,\\
        ~~~\epsilon_{2,f}\qquad &N>N_1+N_{\text{eff}}\,,
    \end{cases}
\end{align}
here $\epsilon_{2,f} \equiv 2\mu-3$ is the asymptotic value of our solution $\epsilon_2$ in Eq.~\eqref{eq:sole2} at the end of SR2 phase. We use $N_1$ to denote the time when the smooth model enters the non-attractor phase $\epsilon_2<-3$, while $N_{\mathrm{eff}}$ is the effective duration of the USR phase, defined as 
\begin{align}\label{def:Neff}
N_{\mathrm{eff}} \equiv-\frac{1}{6} \int_{\epsilon_2<-3} \epsilon_2(N) \mathrm{d}N + \frac{1}{3}\ln \frac{4\mu}{2\mu+3}\,. 
\end{align}
For later convenience, we introduce the shorthand notations $\overline{N}_{\text{eff}}$ and $\Delta N$ to denote the first and second terms in Eq.~\eqref{def:Neff} respectively. $\overline{N}_{\rm{eff}}$ has a transparent interpretation: $6\times\overline{N}_{\rm{eff}}$ is exactly the area of the \textcolor[RGB]{10,190,180}{green}-shaded region in Fig.~\ref{fig:epsilon2_smoothVSsharp}, namely where the smooth model remains in the $\epsilon_2<-3$ regime. The detailed meaning and implications were investigated in Ref. \cite{Pi:2022zxs} for the models ending with $\epsilon_{2,f}=0$, where the peak amplitude is estimated as
$ ( e^{6\overline{N}_{\mathrm
{eff}}} \times 2.61 \times 4)\,\mathcal{P}_{\mathcal{R}}^{
\mathrm
{SR1}}$.
In transient models, $\epsilon_{2,f}$ affects both the location of the dip and the height of the peak. To incorporate these effects, we introduce $\Delta N$, which vanishes for $\epsilon_{2,f}=0$ (equivalently, $\mu=3/2$). With this additional term, a general transient model with non-zero $\epsilon_{2,f}$ can be mapped onto the special case with $\epsilon_{2,f}=0$ such that both the dip position and the peak amplitude are matched exactly. 

\begin{figure}[t!]
\centering
\hspace{-0.5cm}
\includegraphics[width=0.47\textwidth]{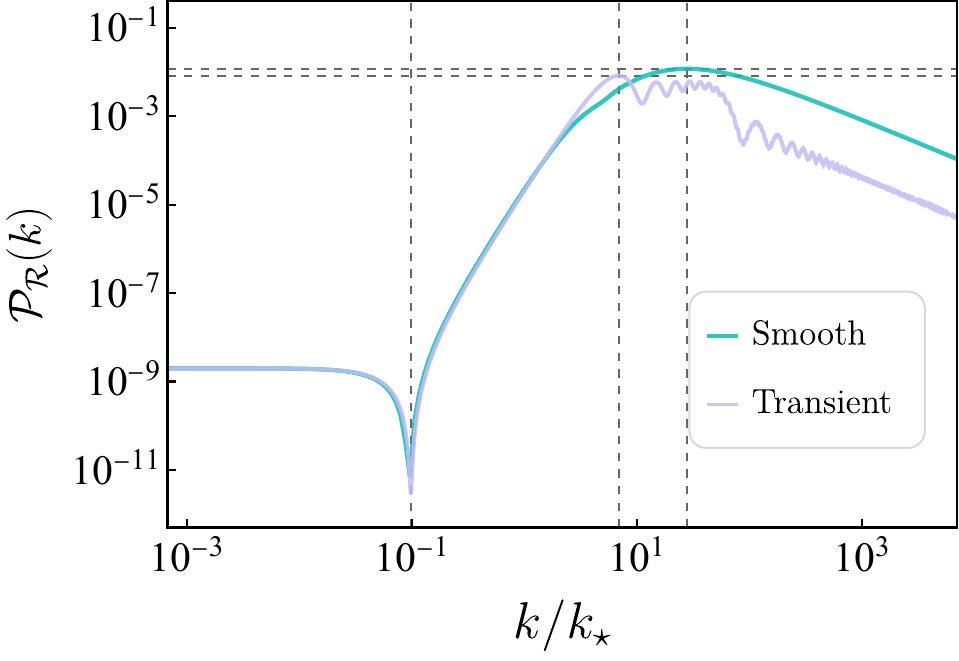}
\caption{The comparison of power spectra between our smooth model (\textcolor[RGB]{10,190,180}{green line}) with $\tau_{\star}=-1$, $\alpha=22.63$ and $\mu=2.0294$, and the transient model (\textcolor[RGB]{197,187,245}{purple line}) given by Eq.~\eqref{eq:sharpmodel}.  Horizontal dashed lines denotes the dip and peak positions, while vertical dashed lines show the amplitude of peaks.}
\label{fig:PowerSpectrum_Compare}
\end{figure}

In Fig.~\ref{fig:PowerSpectrum_Compare}, we compare the power spectrum from our smooth model with that of the transient model, whose $\epsilon_2$ is given by Eq.~\eqref{eq:sharpmodel}. These two models agree perfectly in the IR limit, including the dip position. The peak amplitude differs slightly, by roughly a factor of $\sim1.4$. However, the peak position is noticeably shifted due to the different growth behavior discussed in the previous section. More importantly, the disagreement becomes significant in the UV tail, where the amplitudes differ by factors of $\mathcal{O}(10)$ to $\mathcal{O}(10^2)$. 
The enhanced UV end of the  power spectrum or oscillatory UV features can be easily tested through the associated IGW signals. We leave the detailed investigation of these observational consequences to future work. Finally, for future reference, we also list in Table~\ref{tab:8deltas} several representative cases, beyond the parameter choices considered in the main text, for which the minimum value of $\epsilon_2$ is $-6$. These cases are subject to observational constraints on the PBH abundance.

\vspace{0.5cm}
\noindent\textbf{\emph{Summary}---} In this work, we show, to the best of our knowledge, the first analytically solvable single-field inflation model that allows a smooth SR-USR-SR transition
without introducing any discontinuity in the second slow-roll parameter $\epsilon_2$.  This is realized by a simple ansatz that the effective mass term in the MS equation is a polynomial in time. 
In contrast to previous numerical studies \cite{Ragavendra:2020sop,Cheng:2021lif,Tomberg:2022mkt,Davies:2023hhn,Iacconi:2023ggt,Caravano:2024moy,Jackson:2024aoo,Caravano:2025diq}, the analytical tractability
of our model makes it straightforward to trace both the parameter dependence and the behavior of the power spectrum across different regimes. We further uncover a simple correspondence between our smooth-transition model and the conventional sharp-transition USR scenarios. The two share similar features around the dip and yield comparable peak amplitudes, but they differ significantly in the peak location and in the behavior of the UV tail.

This \textit{letter} focused primarily on the analytical solutions of the model. Many aspects remain to be explored in future works, from both phenomenological and theoretical perspectives. For example, it will be interesting to carry out a detailed analysis of the PBHs abundance and the IGWs signals predicted by our setup. As an analytical expression for the power spectrum is available, we expect to have an analytical expression for IGW spectrum, as is done for monochromatic \cite{Kohri:2018awv,Espinosa:2018eve}, lognormal \cite{Pi:2020otn,Dandoy:2023jot}, and broken power-law \cite{Li:2024lxx} spectra. Other directions worth investigating include non-Gaussianities, as well as stochastic and quantum effects associated with the model. 
All of these issues are closely connected to forthcoming observational probes including micro-lensing surveys \cite{
MACHO:2004slp, 
OGLE:2009crb, 
MOA:2013mrc,DES:2018qoi,Shah:2025eik}, CMB experiments \cite{WMAP:2010qai,
BICEP2:2015xme,
Planck:2018vyg,
LiteBIRD:2023zmo,
SPT-3G:2025vyw,AtacamaCosmologyTelescope:2025blo}, and next-generation gravitational-wave detectors \cite{
Sutton:2008zza,
Kokeyama:2020dkg,
NANOGrav:2023bts,Kawamura:2023fwf,LIGOScientific:2024sgg,ET:2025xjr,TaijiScientific:2021qgx,
TianQin:2020hid}.
Analytical control over full dynamics makes our smooth scenario both a useful theoretical laboratory and a practical framework for connecting small-scale inflationary physics with observations.

\vskip 5pt
\paragraph*{Acknowledgements}

This work is supported by the National Key Research and Development Program of China Grant No. 2021YFC2203004. 
D.C. is supported by the 78th Batch of General Grants of the China Postdoctoral Science Foundation No. 2025M783447.
M.H. is supported by IBS under the project code, IBS-R018-D1. 
S.P. is supported in part by the National Natural Science Foundation of China (NSFC) Grants No. 12475066, No. 12447101 and JSPS KAKENHI No.~24K00624.
J.W. is supported in part by JSPS KAKENHI No.~24K00624 and by Kavli IPMU, which was established by the World Premier International Research Center Initiative (WPI), MEXT, Japan. 
M.Y. is supported by IBS under the project code, IBS-R018-D3, and by JSPS Grant-in-Aid for Scientific Research Number JP23K20843.
Y.Z. is supported by IBS under the project code, IBS-R018-D3.

\bibliography{Refs}
\bibliographystyle{utphys}

\clearpage
\onecolumngrid
\newpage
\appendix
\section{Coefficients of solutions} \label{apdx:coeffi}
In this appendix, we provide the explicit expressions for all coefficients in the analytical solutions presented in the main text.
\\

$\bullet$ \textit{{coefficients of $\epsilon_2(N)$}}:
\\
\\
$\epsilon_2$ satisfies Eq.~\eqref{evol_epsilon2_N}, a first-order ordinary differential equation, together with the boundary condition $\epsilon_2(\tau_\star)=0$, the solution can be fixed as 
\begin{equation}\label{eq:sole2}
\epsilon_2(N)=e^{-N}\,\frac{\mathcal{G}_1(0) \cdot \mathcal{G}_2(N)-\mathcal{G}_2(0) \cdot \mathcal{G}_1(N)}{\mathcal{G}_1(0) \cdot \mathcal{G}_4(N)+\mathcal{G}_2(0) \cdot \mathcal{G}_3(N)}\,,
\end{equation}
where functions $\mathcal{G}_{i}(N)$ are 
\begin{align}\label{eq:funsineta}
\mathcal{G}_1(N) &= \left[e^{N}(3+2\mu)-2q\right] \,\,U\left[\frac{1}{2}-\frac{\alpha}{2q}+\mu,1+2\mu,2 q\,e^{-N}\right]+2(\alpha-q-2q\mu)\,\,U\left[\frac{3}{2}-\frac{\alpha}{2q}+\mu,2+2 \mu,2 q\,e^{-N}\right]\,,\\
\mathcal{G}_2(N)&=4 q\,\,L_{\alpha/2q-\mu-3/2}^{1+2\mu}\left(2 q\, e^{-N}\right)-\left[e^{N}(2\mu+3)-2q\right] ~L_{\alpha/2q-\mu-1/2}^{2 \mu}\left(2 q\,e^{-N}\right)\,,\\
\mathcal{G}_3(N)&=U\left[\frac{1}{2}-\frac{\alpha}{2q}+\mu,1+2\mu,2 q\,e^{-N}\right]\,,\\
\mathcal{G}_4(N)&=L_{{\alpha}/{2q}-1/2-\mu}^{2 \mu}\left(2 q\,e^{-N}\right)\,,
\end{align}
here $U(a,b,z)$ is the confluent Hypergeometric-$U$ function, while $L_n^a(z)$ is the generalised Laguerre polynomial.
\\

\par$\bullet$ \textit{{coefficients of $z(\tau)$}}:
\\
\\
The solution is given by Eq.~\eqref{eq:solution_of_z}, and the coefficients are
\begin{align}
\label{eq:b1b2}
&\mathcal{B}_1
=\frac{\sqrt{2\epsilon_{\mathrm{SR1}}}M_{\rm{pl}}}{H_0 \tau_\star}\,\frac{(\alpha-2 q-2 q^2 ) M_0-(\alpha+q+2\mu q) M_1}{(\alpha+q+2\mu q) M_1 W_0+2 q M_0 W_1}\,,\\
&\mathcal{B}_2
=-\frac{\sqrt{2\epsilon_{\mathrm{SR1}}}M_{\rm{pl}}}{H_0 \tau_\star}\,\frac{(\alpha-2 q-2 q^2 ) W_0+2 q W_1}{(\alpha+q+2\mu q) M_1 W_0+2 q M_0 W_1}\,,
\end{align}
where we have introduced the shorthand notations as
\begin{align}
{W}_n\equiv W_{\frac{\alpha}{2 q}+n, \mu}\left( 2 q \right)\,,\quad\quad
M_n\equiv M_{\frac{\alpha}{2 q}+n,\mu}\left( 2 q \right)\,.
\end{align}
\par$\bullet$ \textit{{coefficients of $v_k(\tau)$}}:
\\
\\
The solution of MS variable $v_k(\tau)$ is given by Eq.~\eqref{eq:solution_of_v}, and the coefficients are
\begin{align}\label{eq:C1C2}
\mathcal{C}_1(k)&= \frac{e^{-i k \tau_\star}}{\sqrt{2}k^{3/2}\tau_\star \mathcal{Y}_0}
\,\bigg\{ \left[-i \frac{\alpha}{\tau_\star}+\left(1+2\mu\right)k_{\mathrm{eff}}\right] \left( -i+k \tau_\star\right)\,\widetilde{M}_1 + \left[  \left( 1+i k \tau_\star\right)\left( 2 k_{\text{eff}}  \left(k+k_{\text{eff}}\right)\tau_\star  +\frac{\alpha}{\tau_\star} \right) -2 i k_{\text{eff}}\right]\, \widetilde{M}_0\bigg\}\,,
\\
\mathcal{C}_2(k)&= \frac{e^{-i k \tau_\star}}{\sqrt{2}k^{3/2}\tau_\star \mathcal{Y}_0}\,\bigg\{ \left[2 i k_{\text{eff}} + \left( 1+i k \tau_\star\right)\left(-\frac{\alpha}{\tau_\star} - 2 k_{\text{eff}}  \left(k+k_{\text{eff}}\right)\tau_\star \right) \right]\, \widetilde{W}_{0}-2 i k_{\mathrm{eff}} \left( 1+i k \tau_\star\right) \widetilde{W}_{1} \bigg\}\,, 
\end{align}
we also defined the shorthand notations as 
\begin{align}
\widetilde{W}_n\equiv W_{i\widetilde{\alpha}+n, \mu}\left( 2 i  k_{\rm{eff}} \tau_\star\right)\,,\quad\quad
\widetilde{M}_n\equiv M_{i\widetilde{\alpha}+n,\mu}\left( 2 i  k_{\rm{eff}} \tau_\star\right)\,,
\end{align}
and the factor $\mathcal{Y}_0$ in the denominator then can be expressed as 
\begin{align}
\mathcal{Y}_0=&2k_{\mathrm{eff}}\,\widetilde{W}_1\,\widetilde{M}_0+\left[-i \frac{\alpha}{\tau_\star}+(1+2\mu)k_{\mathrm{eff}}\right]\,\widetilde{W}_0\,\widetilde{M}_1\,.
\end{align}

\section{ Approximation formulae in different regions} \label{apdx:approx}
First of all, to obtain the final power spectrum Eq.~\eqref{eq:R_exact}, we have used the  asymptotic behaviour of Whittaker functions:
\begin{align}
    &\lim_{z\to0} M_{\kappa,\mu}(z)=z^{\frac{1}{2}+\mu}\,,\\
    &\lim_{z\to0}W_{\kappa,\mu}(z)=z^{\frac{1}{2}-\mu}\frac{\Gamma(2\mu)}{\Gamma(1/2-\kappa+\mu)}+z^{\frac{1}{2}+\mu}\frac{\Gamma(-2\mu)}{\Gamma(1/2-\kappa-\mu)}\,,
\end{align}
in the case $\mu>0$, the dominant contribution arises from the  $z^{1/2-\mu}$ term of the Whittaker-$W$, Eq.~\eqref{eq:R_exact} then reduces to a ratio of Euler Gamma functions.
\\
\\
$\bullet$ \textit{IR region}:
\\
\\
In the IR region $k / k_\star\ll 1$, we first take ${k_{\rm{eff}}}\to iq k_\star$ in the Whittaker functions\footnote{Strictly speaking, we have to perform the expansion while keeping the $ k $-dependence in both the parameters and the arguments of the Whittaker functions, which would lead to more complicated expressions involving derivatives of the Whittaker functions with respect to their parameters.
But we have checked that the approximated results have at most percent-level errors for the parameters chosen in our study, which does not affect our main results and conclusion, so we adopt this approximation for simpler expression. 
Moreover, some software do not provide built-in functions for direct evaluation of the derivatives of the Whittaker functions.} and then perform the expansion around $ k=0 $, which yields a quadratic function, whose coefficients are given as
\begin{align}
a_0&= \frac{1}{2\tau_\star^2}\left[\frac{M_0(\alpha-2q-2q^2)-M_1(\alpha+q+2\mu q)}{2q\,M_0W_1+M_1 W_0(\alpha+q+2\mu q)}\right]^2\,,\\
\frac{a_2}{a_0} &= \tau_{\star}^2 \frac{\left( \alpha +q + 2\mu q \right)M_1 + \left( 2q^2 -2q -\alpha \right) M_0}{\left( \alpha +q + 2\mu q \right)M_1 +\left( 2q^2 +2q -\alpha \right) M_0} ~.
\end{align}
\\
\\
$\bullet$ \textit{Growth region}:
\\
\\
For the $k^4$-growth region, the key quantity is the dip position which can be determined from the expressions for $a_i$ above. To analyse the other growth behavior, we need the power spectrum evaluated at $k=q$. At first glance, this may appear problematic, since $k_{\rm{eff}}\to0$ and then the Whittaker-function parameter $\widetilde{\alpha}$ is divergent. To evaluate it properly, we need to use the following asymptotic formula \cite{NIST:DLMF},
\begin{align}
   \lim_{\kappa\to\infty}M_{\kappa,\mu}(x)&=\kappa^{-\mu}\sqrt{x}\,\Gamma(2\mu+1)\,J_{2\mu}(2\sqrt{x\kappa})\,, \\
   \lim_{\kappa\to\infty}W_{\kappa,\mu}(x)&=\sqrt{x}\,\Gamma(\kappa+1/2)\Big[\sin\pi(\kappa-\mu)\,J_{2\mu}(2\sqrt{x\kappa})-\cos\pi(\kappa-\mu)\,Y_{2\mu}(2\sqrt{x\kappa})\Big]\,,
\end{align}
here $J_{\mu}(x)$ and $Y_{\mu}(x)$ are the Bessel functions of the first and second kinds, respectively. Then we can get the power spectrum 
\begin{align}
    \label{eq:Psig}
    &\mathcal{P}_{\mathcal{R}}(q k_\star)\nonumber\\
    &=\frac{(2q)^{2\mu-1}}{|4\pi\tau_\star\mathcal{B}_1|^2}\,\Gamma\left(\frac{1}{2}-\frac{\alpha}{2q}+\mu\right)^2 \bigg| \Big((3+2\mu)(i-q)-2iq^2 \Big){}_0\mathbf{F}_1\left(;2 \mu+1;-\alpha \right) - 2\alpha\left(i-q\right) \, {}_0\mathbf{{F}}_1\left(;2 \mu+2;-\alpha \right)\bigg|^2\,,
\end{align}
where ${}_0\mathbf{F}_1\left(;a;z\right) $ is a regularised confluent Hypergeometric function.
\\
\\
$\bullet$ \textit{UV region}:
\\
\\
In the region $k / k_\star\gg1$, the parameter appearing in the special functions becomes $\widetilde{\alpha}\to 0$. This makes it straightforward to expand the power spectrum in the UV region, yielding
\begin{align}
    \mathcal{C}_1 (k)\sim \frac{1}{\sqrt{2k}}\,,
\end{align}
and $k_{\rm{eff}}\to k$ in the Gamma functions, then we can readily obtain the final approximation Eq.~\eqref{eq:PUV}.

\section{The approximation of the scale factor} \label{apdx:dS}

In this appendix, we justify the approximation $a=-1/{H_0 \tau}\,, N=-\ln\left(\tau/\tau_\star\right)$ in our analysis. 
By definition, the $e$-fold number is 
\begin{align}
    N\equiv \ln \frac{a}{a_0} ~,
\end{align}
which allows us to write down 
\begin{align}
    a= a_0 \, e^N ~, ~\dd N = H \dd t ~. 
\end{align}
Therefore, the first slow-roll parameter is given by 
\begin{align}
    \epsilon_1 \equiv - \frac{\dot{H}}{H^2} = - \frac{\dd \ln H}{\dd N} ~, 
\end{align}
which can be solved as 
\begin{align}
    H = H_0 \exp \left( -\int \epsilon_1 \dd N \right) ~.
\end{align}
As a result, the conformal time can be expressed as 
\begin{align}
    \tau = \int \frac{\dd t}{a} = \int \frac{\dd N}{a H} &=\int \frac{\dd N}{a H_0} \exp \left( \int^N \epsilon_1 \dd N' \right)=\int \frac{\dd N}{a_0 H_0} e^{-N} \sum_{n=0} \frac{1}{n!} \left( \int^N \epsilon_1 \dd N' \right)^n ~. 
\end{align}
For the exact de Sitter case, $ \epsilon_1 =0 $ such that 
\begin{align}
    \tau = -\frac{1}{a H_0} ~. 
\end{align}
Slow-roll correction comes from the higher order terms in the Taylor expansion. 
These terms can be safely dropped as long as parameters are properly chosen such that 
\begin{align}
    \left| \int \epsilon_1 \dd N \right| \ll 1 ~,
\end{align}
which is exactly the case in the main text.

\section{Other choices of parameters} \label{apdx:table}

\begin{table}[h]
\centering
\renewcommand{\arraystretch}{1.4}
\setlength{\tabcolsep}{11pt}
\begin{tabular}{c|c|c|c|c}
\hline
\hline
$\mu$  & $\alpha$ & $N_1$ & $N_{\mathrm{eff}}$ & Max $\mathcal{P}_{\mathcal{R}}$ \\
\hline
\hline
{2.2407} & 25.860 & 0.720 & 1.90 & $1.92\times 10^{-3}$ \\ \hline
{2.1681} & 24.730 & 0.735 & 2.00 & $3.59\times 10^{-3}$ \\ \hline
{2.0976} & 23.650 & 0.750 & 2.06 & $5.45\times 10^{-3}$ \\ \hline
{2.0294} & 22.630 & 0.766 & 2.19 & $1.18\times 10^{-2}$ \\ \hline
{1.9633} & 21.660 & 0.781 & 2.34 & $3.07\times 10^{-2}$ \\ \hline
{1.8991} & 20.735 & 0.797 & 2.47 & $7.18\times 10^{-2}$ \\ \hline
{1.8368} & 19.855 & 0.814 & 2.63 & $1.84\times 10^{-1}$ \\ \hline
{1.7764} & 19.018 & 0.830 & 2.82 & $6.24\times 10^{-1}$ \\ \hline
\end{tabular}
\caption{Peak amplitudes for different parameter choices. In order to make connection with USR, the parameters are fine-tuned such that the minimum value of $\epsilon_2$ is $-6$. Here $\mu$ and $\alpha$ are parameters in our smooth transition model, while $N_1$ and $N_{\mathrm{eff}}$ are parameters in the sharp transition counterpart. }
\label{tab:8deltas}
\end{table}

Tab.~\ref{tab:8deltas} summarizes several representative examples in which $\epsilon_2$ reaches the critical value $-6$, together with the corresponding peak amplitude $\max \mathcal{P}_{\mathcal{R}}$. Demanding that PBHs are not overproduced sets an upper limit on the peak amplitudes, which in turn imposes a lower bound on the parameter $\mu$ in our model. As shown in the table, values $\mu \gtrsim 2$ are preferred.

\end{document}